\documentstyle[budapest1]{article}  
\frompage{000} \topage{000}                                              
\def\rightmark{pQCD Results on pion production} 
\def\leftmark{G.G.  Barnaf{\"o}ldi et al.}
\def\la{\langle}
\def\ra{\rangle}
\def\beq{\begin{equation}}
\def\eeq{\end{equation}}
\def\be{\begin{eqnarray}}
\def\ee{\end{eqnarray}}
\def\hs{\hat{s}}
\def\htm{\hat{t}}
\def\hu{\hat{u}}

\def\k2av{\la k_T^2\ra}
\def\lesssim{\leq}
\def\gtrsim{\geq}
\newcommand{\f}[2]{\frac{#1}{#2}}
\newcommand{\dd}{ {\textrm d}}

\title{Perturbative QCD Results on Pion Production in \\ 
$pp$, $pA$ and $AA$ Collisions} 
\authors{
{G.G.   Barnaf{\"o}ldi$^{1,2}$, P.  L{\'e}vai$^1$, G.  Papp$^{3}$, 
G.  Fai$^{4}$ and Y.  Zhang$^4$}\\[2.812mm]
{\normalsize
\hspace*{-8pt}$^1$ KFKI Research Institute for Particle and Nuclear Physics, \\ 
P.O. Box 49, Budapest, 1525, Hungary\\[0.2ex] 
\hspace*{-8pt}$^2$ Laboratory for Information Technology, E\"otv\"os
University,\\
P{\'a}zm{\'a}ny P{\'e}ter s{\'e}t{\'a}ny 1/A, Budapest, 1117 Hungary\\[0.2ex]
\hspace*{-8pt}$^3$ HAS Research Group for Theoretical Physics,
E\"otv\"os University \\
P{\'a}zm{\'a}ny P{\'e}ter s{\'e}t{\'a}ny 1/A, Budapest, 1117 Hungary\\[0.2ex]
\hspace*{-8pt}$^4$ Center for Nuclear Research, Department of Physics,\\ 
Kent State University, Kent, OH 44242, USA
}}

\abstract{We summarize new pQCD results on pion production in proton-proton
        ($pp$), proton-nucleus ($pA$) and nucleus-nucleus ($AA$) collisions.
        Our calculation introduces intrinsic parton transverse momentum 
        ($k_T$) and  is performed  effectively at next-to-leading order 
        (NLO), applying a $K$ factor extracted for jet events. 
        Two different factorization scales, 
        $Q=p_{T,jet}/2 $ and $p_{T,jet}$ are used.
        Experimental data in $pA$ collisions imply a preference for
        the latter choice at NLO level. We display our results at CERN SPS 
        for $AA$ collisions.} 

\begin{document}
 
\maketitle
\setcounter{page}{1}


\section{Introduction}
\label{sec_intro}

Recent RHIC and CERN experiments require solid theoretical baseline 
calculations, able to reproduce proton-proton,
proton-nucleus and the measured nucleus-nucleus data accurately
enough to find possible deviations due to new collective phenomena or
to a new state of the matter.
In this paper pion production is calculated in a pQCD-improved parton 
model~\cite{FF95}. 
Recently we have performed a leading order (LO) calculation and
displayed the results of a systematic analysis of 
$pp$ ($p{\overline p}$), $pA$ and $AA$ data on pion production
\cite{zfpbl02}. The key points were the introduction of
the intrinsic transverse momentum distribution of the partons
and the folding of nuclear multiscattering effects
into the pQCD calculations.
We have also calculated pion production in next-to-leading 
order (NLO)~\cite{plf00}. However, the NLO cross sections were estimated
by applying an oversimplified constant $K$ factor ($K=2$) to
the LO cross sections.
Since the determination of nuclear effects (e.g. the strength of
nuclear multiscattering) and other collective effects require great
precision (better than $10-20 \ \%$), we need to repeat the earlier
calculation using appropriate NLO cross sections.
In Section 2 we summarize our results on energy and
momentum dependent $K$ factors.

In standard calculations for $pp$ collisions the factorization
scale is typically chosen in the range
$p_T/3 \leq Q \leq 2p_T$.
We find that at the NLO level the reproduction of
nuclear effects in $pA$ collisions 
(specifically the Cronin effect~\cite{cronin75,antreasyan79})
strongly depends on the choice of the factorization scale. 
We discuss this issue in Section 3.


\begin{figure}[htb] 
\vspace*{-1.1cm}
                 \insertplot{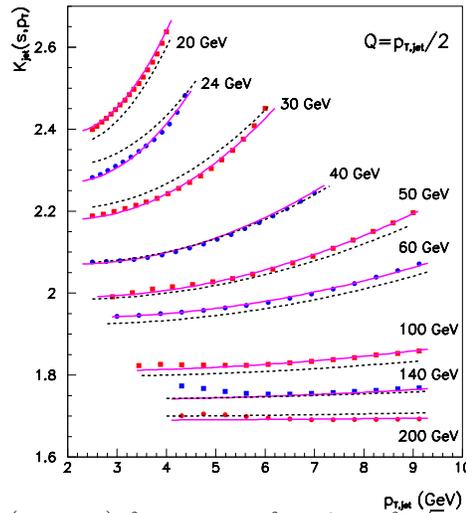}
\vspace*{-0.9cm}
\caption[]{The $K_{jet}(s, p_{T,jet})$ factor as a function of $ \sqrt s $ 
and $p_{T,jet}$ at scale $Q=p_{T,jet}/2$. Solid lines guide the eye
through the calculated points and the fitted parabolae of 
eq.~(\ref{kjet1}) are indicated by dashed lines~\cite{bflpz01}.  }
\label{fig1} 
\end{figure}

\section{$K$ factor for NLO jet-production in $pp$ collision }

In a previous paper~\cite{bflpz01} we have extracted a $K$ factor 
for jet production in $pp$ collisions. The scale in the parton distribution
functions (PDFs) was chosen to be $Q=p_{T,jet}/2$ 
for jets with jet-cone angle $R=1$ and separation $R_{sep}=2R$. 
An energy and transverse momentum dependent correction factor, 
$K_{jet}(s, p_{T,jet})$, was then obtained as
\begin{equation} \label{kdef}
\frac{\dd \sigma^{NLO}}{\dd {\hat t}} = K_{jet} \left( s , p_{T,jet} \right) \,\, 
\frac{ \dd \sigma^{Born}}{ \dd {\hat t}} \,\, ,
\label{Kfac}
\end{equation}
where the cross sections are understood at the jet level 
(see eq.~(\ref{hadX})).

Full NLO calculations were performed by the 
Ellis\,--\,Kunszt\,--\,Soper \mbox{(EKS)~\cite{eks1,eks2}} and the 
Aversa\,--\,Chiappetta\,--\,Greco\,--\,Guillet (ACGG)~\cite{acgg01} groups,
and are based on the matrix elements published in Ref.~\cite{ellis01}.
We used a public Fortran code from the EKS group~\cite{eks03} 
with MRST (central gluon) PDFs~\cite{kimber00}
to calculate the Born and the NLO contributions.
The extracted $K$ factor of eq.~(\ref{Kfac}) 
is shown in \mbox{Fig. 1} in the energy range 
$20$ GeV $\leq \sqrt s \leq 200 $ GeV. 

For practical purposes, 
the $K$ factors were parameterized 
(within 2-4 \% precision) as 
\begin{equation}
  K_{jet}(s , p_{T,jet} )  =
   1.6 + \frac{20.}{\sqrt s } - \frac{24.}{(\sqrt s -10.)^2} \, p_{T,jet}  + 
\frac{6.}{(\sqrt s -10.)^2}      \, p_{T,jet}^2 \,\,\, ,
\label{kjet1}
\end{equation}
where $p_{T,jet}$ and $\sqrt{s}$ are in GeV and 
the constants are understood with their appropriate units.

In this paper we plan to use a different scale in the PDFs (see Section 3), 
namely $Q=p_{T,jet}$. Thus we repeat our calculation of the 
jet-level $K$ factor
at this new scale, using $R=1$. The results are displayed in Fig. 2.
\begin{figure}[htb]
\vspace*{-1.1cm}
                 \insertplot{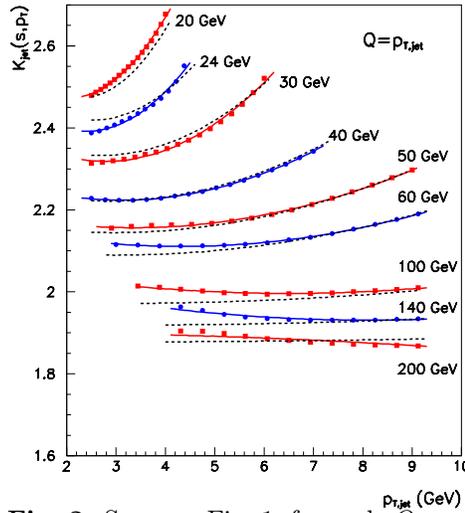}
\vspace*{-0.9cm}
\label{fig2}
\caption[]{Same as Fig.~\ref{fig1}, for scale $Q=p_{T,jet}$.}
\end{figure}

\noindent
These new results may be again parameterized quadratically with good precision:
\begin{equation}
  K_{jet}(s ,p_{T,jet})  =
   1.79 + \frac{20.}{\sqrt s } - \frac{45.}{(\sqrt s -7.)^2} \, p_{T,jet}  
+ \frac{7.}{(\sqrt s -9.)^2}     \, p_{T,jet}^2   \,\,\, .    
\label{kjet2}
\end{equation}

Figures \ref{fig1} and \ref{fig2}  demonstrate that increasing the scale
by a factor of 2, the jet-level $K$ factor is increased by 
$\sim \ 10-15\  \%$, 
maintaining the characteristic dependence on $s$ and $p_T$.
At RHIC energies ($\sqrt{s}=200$ AGeV)
the $K$ factor is almost constant with values
$\sim 1.7$ (at $Q=p_{T,jet}/2$) and $\sim 1.9$ (at $Q=p_{T,jet}$).


\section{Pion production from CERN to Tevatron energies}
\label{sec_pi}

Here, we summarize 
the method used to calculate pion production 
in a pQCD-improved parton model, following our previous work
in LO, see Ref.~\cite{zfpbl02}.  In Subsection \ref{sec_kT}
we introduce the intrinsic transverse momentum distribution for
partons and fit the width of the applied Gaussian  
distribution to available $pp$ (or $p \bar{p}$) data in the c.m. energy 
range 20 GeV $\lesssim \sqrt{s} \lesssim$ 1.8 TeV. 
A large collection of pion and charged hadron production data 
can be utilized in this energy range to extract information about 
the intrinsic transverse momentum distribution,
e.g the energy dependence of the Gaussian width. In Subsection \ref{sec_nu}
the Cronin effect~\cite{cronin75,antreasyan79} is discussed in $pA$ collisions. 
Subsection \ref{sec_AApi} deals with hard pion production in $AA$ 
collisions at CERN SPS energies.

\subsection{Parton model calculations with intrinsic transverse momentum}
\label{sec_kT}     

In $pp$ collision we describe the
invariant cross section for pion production in a 
pQCD-improved parton model on the basis 
of the factorization theorem~\cite{FF95}: 
\begin{eqnarray}
 E_{\pi}\f{\dd \sigma_{\pi}^{pp}}{\dd ^3p} &=&
 \sum_{abcd} \int\! \dd x_a \, \dd x_b \, \dd z_c \ f_{a/p}(x_a,Q^2)\
 f_{b/p}(x_b,Q^2)\ \times \  \nonumber \\
 \  & \times & \ \left[ K_{jet}(s ,p_{T,jet})  
 \, \, \f{\dd \sigma}{\dd \htm}^{ab \rightarrow cd}\,
 \right]  \frac{D_{\pi/c}(z_c, Q'^2)}{\pi z_c^2} 
 \,\, \hs \,\, \delta(\hs+\htm+\hu)\ . 
\label{hadX}
\end{eqnarray}
\noindent
Here  $f_{a/p}(x_a,Q^2)$ and $f_{b/p}(x_b,Q^2)$ are the 
LO/NLO parton distribution
functions  for the colliding partons $a$ and $b$ within the interacting 
protons as functions of momentum fraction $x$, at scale $Q$.
$\dd \sigma/ \dd\htm$ is the hard scattering cross section of the
partonic subprocess $ab \to cd$ in LO (Born term)
and the $K_{jet}$ factor is applied
to include NLO contributions.
In eq.~(\ref{hadX}) we use the convention that the parton-level 
Mandelstam variables are indicated by a `hat' (like $\htm$ above). 
The LO/NLO fragmentation function (FF),
$D_{\pi/c}(z_c, Q'^2)$ gives the probability for parton $c$ to fragment 
into $\pi$ with momentum fraction $z_c$ and fragmentation scale $Q'$. 
For the results presented here we fix $Q' = p_T/2$ and use
the KKP parametrization \cite{KKP}.

 For LO calculations, $K_{jet}( s, p_{T,jet}) \equiv 1$.
As discussed in Section 2, in NLO calculations the Born
partonic cross section is multiplied by $K_{jet}( s, p_{T,jet})$ inside
the integral of eq.~(\ref{hadX}). 
For LO calculation the GRV~\cite{GRV92}, for NLO
the MRST~\cite{kimber00} PDF sets were applied 
at scales $Q=p_{T,jet}/2$ and $Q=p_{T,jet}$,  respectively, where
$p_{T,jet}=p_T/z_c$. 

In a phenomenological approach, eq.~(\ref{hadX}) can be generalized to 
incorporate intrinsic transverse momentum by using a product assumption and
extending each integral over the parton distribution functions to 
$k_T$-space~\cite{Wang9798,Wong98},
\beq
\label{ktbroad}
\dd x \ f_{a/p}(x,Q^2) \rightarrow \dd x 
\ \dd ^2\!k_T\ g({\vec k}_T) \  f_{a/p}(x,Q^2) \ ,
\eeq
where $g({\vec k}_T)$ is the intrinsic transverse momentum distribution
of the relevant parton in the proton. We follow this 
approach in the present work, choosing $g({\vec k}_T)$ to be a Gaussian:
\beq
\label{kTgauss}
g({\vec k}_T) \ = \f{1}{\pi \la k^2_T \ra}
        e^{-{k^2_T}/{\la k^2_T \ra}}    \,\,\, ,
\eeq
with $\langle k_T^2 \rangle$ being the 2-dimensional width of the $k_T$
distribution. Increasing the value of parameter
$\langle k_T^2 \rangle$, 
the particle production also increases in the 
transverse momentum window $2$ GeV $\leq p_T \leq 6 $ GeV.

\begin{figure}
\vspace*{-1.1cm}
                 \insertplot{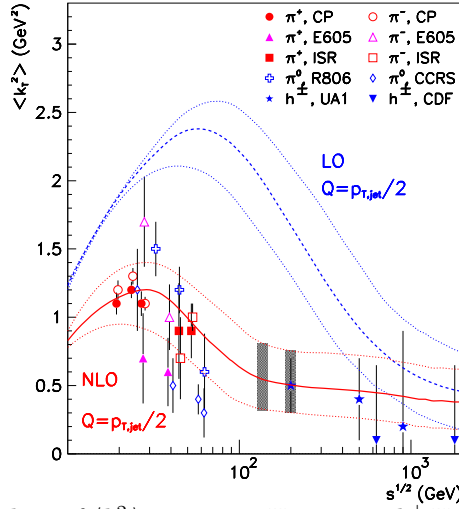}
\vspace*{-0.9cm}
\caption[]{
\label{figure3}
Best fit values of $\k2av$ in $pp \rightarrow \pi X$ or $ p \bar{p} 
\rightarrow h^{\pm} X$ 
reactions at LO ( dashed line) and NLO (solid line) levels
with factorization scale $Q=p_{T,jet}/2$, both with errors (dotted lines). 
Due to the overlap at the same energy, some  
points with error bars have been slightly shifted for better visibility.
Thin lines represent the error estimate of the curves, while the thick bars
are for RHIC energies in NLO.}
\end{figure}

\begin{figure}
\vspace*{-1.1cm}
                 \insertplot{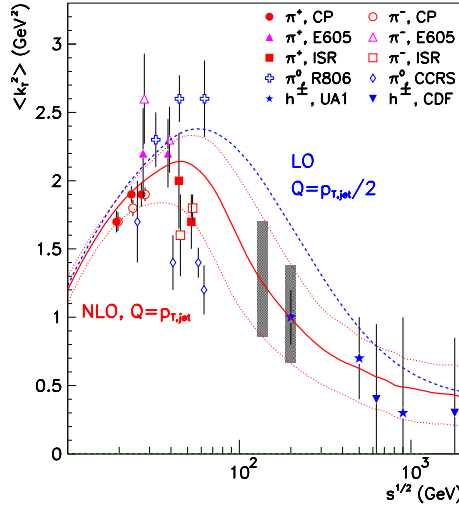}
\vspace*{-0.9cm}
\caption[]{\label{figure4} 
Same as Fig.~\ref{figure3} for scale $Q=p_{T,jet}$ at NLO level compared to
the LO results at $Q=p_{T,jet}/2$ from Fig.~\ref{figure3}.}
\end{figure}

In Figs.~\ref{figure3} and~\ref{figure4}, we present the best fit 
values of $\k2av$, calculated in pion and charged hadron 
($h^{\pm}$) production in $pp$ and $p\bar{p}$ collisions at
scales $Q=p_{T,jet}/2$ and $Q=p_{T,jet}$, respectively, using several 
independent  experimental data 
\cite{antreasyan79,R806,CCRS,E605pi,E605pib,ISR,UA1arn,UA1alb,UA1boc,CDF}.
As expected, the obtained value of $\k2av$ at NLO level  is less 
than at LO level if $Q=p_{T,jet}/2$
is used in both calculations (see Fig.~3).
However, this difference in $\k2av$ 
almost disappears changing the NLO scale to
$Q=p_{T,jet}$, but keeping the LO scale at $Q=p_{T,jet}/2$ (see Fig.~4). 

\begin{figure}[htb]
\vspace*{-1.1cm}
                 \insertplot{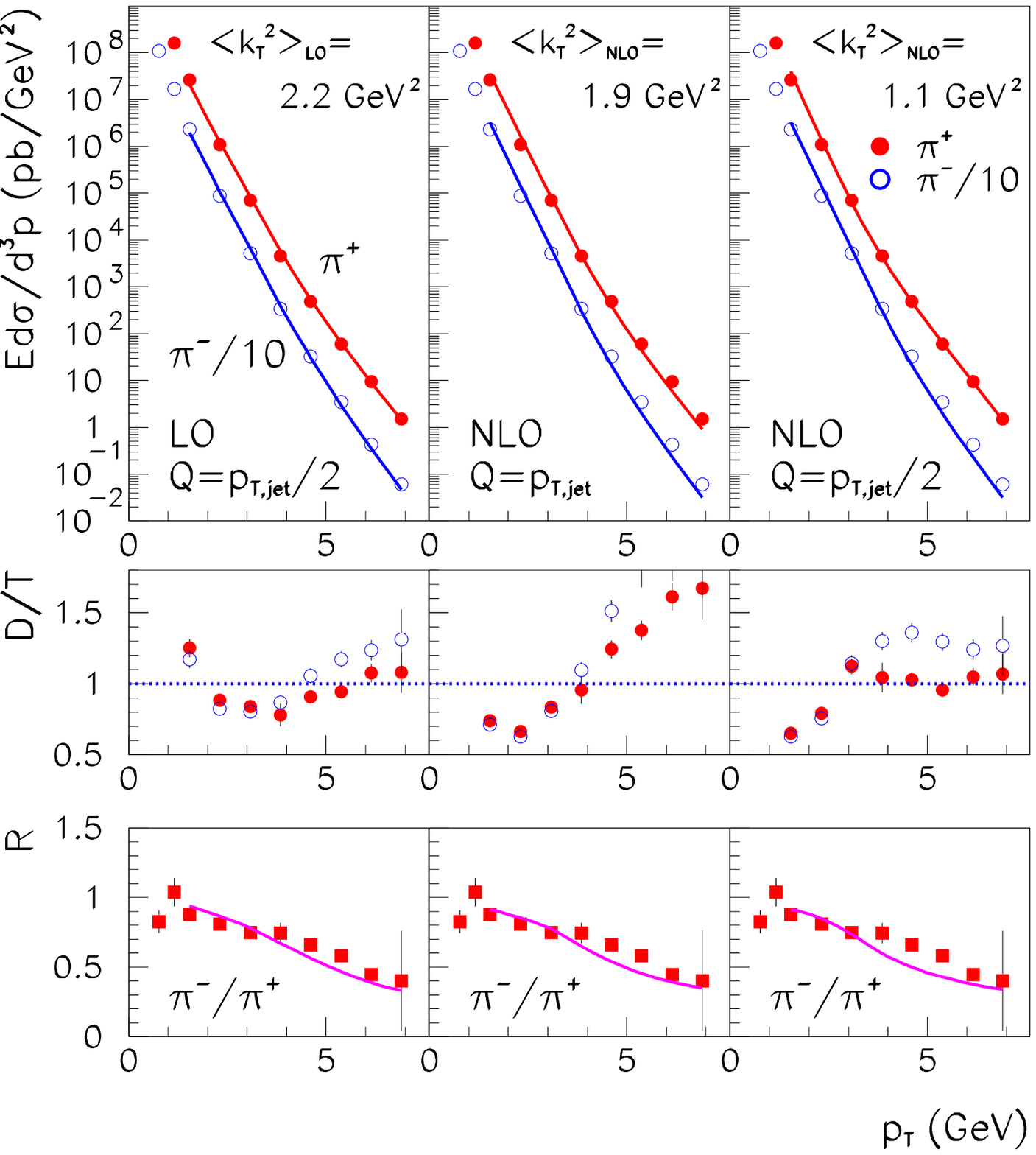}
\vspace*{-1.1cm}
\caption[]{
\label{figurepi} LO and NLO calculations at different scales (see text).
Top panel: invariant cross section of $\pi^+$ and $\pi^-$ production from $pp$
collisions (Negative pion data are divided by 
10 for better visibility). 
Central panel: data/theory ratios, $D/T$.
Lower panel: $\pi^-/\pi^+$ ratios as functions of transverse 
momentum at c.m. energies $\sqrt{s} =$27.4 GeV. Data are from 
\cite{antreasyan79}.}
\end{figure}

As an example to illustrate the degree of accuracy of the description in 
the $pp$ sector, Fig.~\ref{figurepi} compares calculated $\pi^+$ and 
$\pi^-$ spectra and $\pi^-/\pi^+$ ratios to the data~\cite{antreasyan79} at 
the c.m. energy $\sqrt{s} =$ 27.4 GeV, respectively, obtained with the 
values of $\k2av$ (the different scales are indicated in the top panel).
We find that the data/theory and $\pi^-/\pi^+$ ratios are well 
reproduced for 2 GeV $\lesssim p_T \lesssim$ 6~GeV. 
Based on this 
and similar examples, we believe that hard pion production in $pp$ 
collisions is reasonably under control at the present level of calculation.

 
\subsection{Hard pions from $pA$ collisions -- Cronin effect}
\label{sec_nu}

Interest in the nuclear dependence of hard particle production was 
motivated by the discovery of the Cronin 
effect~\cite{cronin75,antreasyan79}:
it was found experimentally that in the transverse momentum window  
$2$ GeV $ \leq p_T \leq 6 $ GeV 
more particles are produced than it was
expected from a simple scaling of the $pp$ data. 
As discussed in the introduction, 
reproducing the Cronin effect could be the key point -- at least in the sense 
of scale fixing -- in our calculations. 

In our model 
there is an extra contribution to the basic Gaussian width 
of the intrinsic parton transverse momentum distribution
due to the nuclear environment. This extra width
can be related to the number of nucleon-nucleon (NN) collisions in 
the medium. 
To characterize the $\k2av$ enhancement, we write the modified width of the 
parton transverse momentum distribution of
the incoming proton as
\beq
\label{ktbroadpA}
\k2av_{pA} = \k2av_{pp} + C \cdot h_{pA}(b) \ . 
\eeq
Here $\k2av_{pp}$ is the width of the transverse momentum distribution 
of partons in $pp$ collisions from Subsection~\ref{sec_kT} (also denoted 
simply by $\k2av$), $h_{pA}(b)$ describes the number of effective
NN collision at impact parameter $b$, each of which imparts an 
average increase in the transverse momentum width denoted by $C$. 
In $pA$ reactions, one parton from the projectile proton and
another one from the target nucleus participate in 
the hard collision. We apply $\k2av_{pA}$ for the projectile parton, which 
incorporates the additional NN collisions indicated in eq.~(\ref{ktbroadpA}).
For the target parton the original $\k2av_{pp}$ is used from  
Fig. \ref{figure3} or Fig. \ref{figure4}.

\begin{figure}
\vspace*{-1.1cm}
                 \insertplot{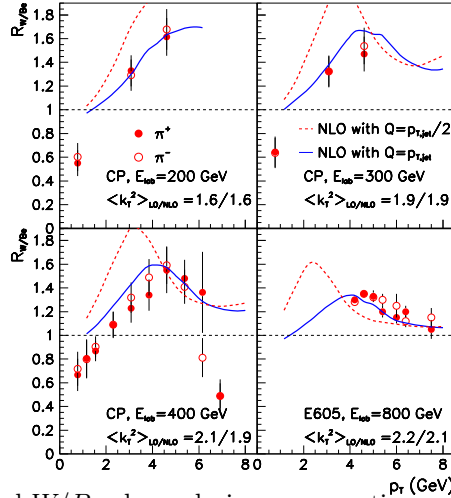}

\vspace*{-1.1cm}
\caption[]{
\label{figurecron}
Normalized $W/Be$ charged pion cross section ratios at different 
energies: $\sqrt s= 19.4; 23.8; 27.4;$ and 
$38.8$ GeV~\cite{antreasyan79,E605pi,E605pib,brown96}. 
The calculations are
at NLO level with scale $Q=p_{T,jet}/2$ (dashed lines) and with 
$Q=p_{T,jet}$ (solid line). The deviation from unity represents 
the Cronin enhancement.}
\end{figure}
The effectivity function $h_{pA}(b)$ can be written in terms of the 
number of collisions suffered by the incoming proton in the target 
nucleus, $\nu_A(b) = \sigma_{NN} t_A(b)$, where $\sigma_{NN}$ is
the inelastic nucleon-nucleon cross section:
\begin{equation}
  h_{pA}(b) = \left\{ \begin{array}{cc}
                \nu_A(b)-1 & \nu_A(b) < \nu_{max} \\
                \nu_{max}-1 & \mbox{otherwise} \\
        \end{array} \right.\ .
\end{equation}

The value $\nu_{max}= \infty$ corresponds to the case where all possible 
semihard
collisions contribute to the broadening. Requiring independence of target
we found that 
for realistic nuclei the maximum number of semihard collisions is 
$2 \leq \nu_{max} \leq 5$.

According to the Glauber picture, the hard pion production cross section 
from $pA$ reactions can be written as an integral over impact parameter $b$:
\beq
\label{pAX}
  E_{\pi}\f{\dd \sigma_{\pi}^{pA}}{ \dd ^3p} =
    \int \dd ^2b \,\, t_A(b)\,\, E_{\pi} \, 
    \f{\dd \sigma_{\pi}^{pp}(\k2av_{pA},\k2av_{pp})}
{\dd ^3p}  
\,\,\, ,
\eeq
where the proton-proton cross section on the right hand side represents 
the cross section from eq.~(\ref{hadX}) with the transverse momentum extension 
as given by eq.-s (\ref{ktbroad}) and (\ref{kTgauss}). Here 
$t_A(b) = \int \dd z \, \rho(b,z)$ is the nuclear thickness function 
(in terms of the density distribution $\rho$) normalized as 
$\int \dd ^2b \, t_A(b) = A$. Furthermore, the PDFs are modified in the 
nuclear environment (``shadowing'')~\cite{wang91,eskola99}. This effect and
isospin asymmetry are taken into account on average using a scale 
independent parameterization of the shadowing function $S_{a/A}(x)$ 
adopted from Ref.~\cite{wang91}:
\begin{equation}
f_{a/A}(x,Q^2) = 
S_{a/A}(x) \left[\frac{Z}{A} f_{a/p}(x,Q^2) + \left(1-\frac{Z}{A}\right)
  f_{a/n}(x,Q^2) \right]   \,\,\,\,  ,
\label{shadow}
\end{equation}
where $f_{a/n}(x,Q^2)$ is the PDF for the neutron and Z 
is the number of protons.

The Cronin enhancement is presented in Fig.~\ref{figurecron} by the normalized 
(with mass number $A$) $W/Be$ charged pion cross section 
ratios together with the data at several energies for $\pi^+$ (dots) 
and $\pi^-$ (open circles)~\cite{antreasyan79,E605pi,E605pib,brown96}. 
In our NLO calculations we used the
previously fixed $\k2av$ values at
scales $Q=p_{T,jet}/2$ (dashed lines) and $Q=p_{T,jet}$ 
(solid lines). Here the Cronin parameter is $C=0.35$ GeV$^2$ and 
$\nu_{max} = 3$. 

In the absence of the Cronin effect, these ratios would be  
identically 1.  The significant deviation of the data from unity is 
therefore a clear confirmation of the nuclear enhancement in the 2 GeV 
$\lesssim p_T \lesssim$ 6 GeV
transverse momentum window. At low $p_T$
the ratio is smaller than unity, indicating absorption effects.

It is clear that the height of the 
Cronin peak depends on the extra $k_T$ broadening
(quantitatively on the product $C \cdot  h_{pA}(b) $), 
and Fig.~\ref{figurecron} tells us that
the location of the peak depends on the value of $\k2av_{pp}$. 
Using our previously fixed parameters, NLO calculations at $Q=p_{T,jet}$ 
(solid lines) give the best fit to the experimental data 
(overlapping with LO curves at $Q=p_{T,jet}/2$, which curves not shown in 
Fig.~\ref{figurecron}, see Ref.~\cite{zfpbl02}). 
In both cases the values of $\k2av_{pp}$ are very much similar
and $\k2av_{pp}\approx 2 \ {\rm GeV}^2$. This characteristic value
of $\k2av_{pp}$ is in good agreement with the experimental 
results of Ref.~\cite{Corcor}.
However, NLO calculations at \mbox{$Q=p_{T,jet}/2$} (dashed lines) with
much smaller $\k2av_{pp}$ (see Fig.~\ref{figure3})
give peaks shifted to 
 unphysically low values \mbox{of $p_T$.} The sensitivity of the Cronin 
effect to the scale at the NLO level dictates our choice of scale 
in the present work.


\subsection{Nucleus-nucleus collisions}
\label{sec_AApi}

In nucleus-nucleus reactions, where both partons entering the hard 
collision originate in nucleons with additional semi-hard collisions, 
we do not need additional parameters, apply the Cronin
enhanced width~(\ref{ktbroadpA}) for both initial partons. Thus,
\begin{equation}
\label{ABX}
  E_{\pi}\f{\dd \sigma_{\pi}^{AB}}{\dd ^3p} =
       \int \dd ^2b \, \dd ^2r \,\, t_A(r) \,\, t_B(|\vec b - \vec r|)  
  \ E_{\pi} \, \f{\dd \sigma_{\pi}^{pp}(\k2av_{pA},\k2av_{pB})}{\dd ^3p}  
\,\,\, .
\end{equation}
For CERN energies we follow Fig.~\ref{figure4} 
and fix the values of $\k2av_{pp}=$ 1.6~GeV$^2$ 
($E_{beam}=158$ AGeV) and 1.7~GeV$^2$ ($E_{beam}=200$ AGeV)
like in Ref.~\cite{zfpbl02} for LO calculations. 
We keep $Q=p_{T,jet}$ and the Cronin parameters
$\nu_m = 3$,  $C = 0.35$~GeV$^2$. 

\begin{figure}
\vspace*{-1.1cm}
                 \insertplot{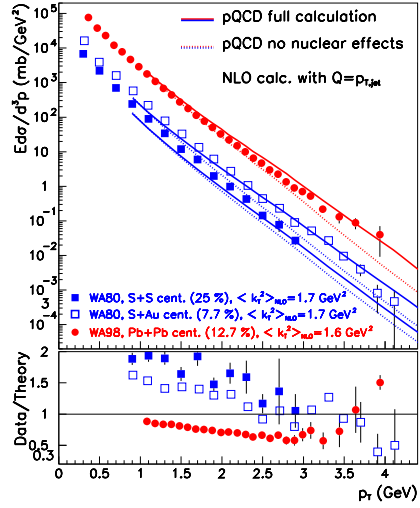}
\vspace*{-0.9cm}
\caption[]{
\label{figure7}
Neutral pion production compared to data from SPS experiments 
WA80 and WA98 for central heavy ion collisions. Top panel shows 
the data and the calculated NLO 
invariant cross sections with the nuclear effects turned off
(dotted lines) and for the full calculation (solid lines).
Bottom panel displays  the data/theory ratios 
for the full calculation.}
\end{figure}

In Fig.~\ref{figure7} the results of our calculation
for $\pi^0$ transverse momentum spectra are 
compared to the WA80~\cite{WA80} and WA98~\cite{WA98} data on
central $S+S$, $S+Au$, and $Pb+Pb$ collisions. 
Top panel shows the spectra, bottom panel displays the data/theory 
ratios. Dotted lines represent the 
results of the pQCD calculation with nuclear effects (shadowing and 
multiscattering) turned off, while solid lines correspond to the full 
calculation. 
(To test the effect of shadowing alone, we turned it off; the
modification resulted in a $\lesssim$ 10\% downward shift 
in the spectra.)  
These calculations were performed down to $p_T=1$ GeV; however,
since pQCD is a theory of hard particle production, it is not 
expected to describe the data below \mbox{$p_T \approx$ 2 GeV.} 

In $S + S$ and $S+Au$ collisions the pQCD results reproduce the data at 
\mbox{$p_T \gtrsim  2.5-3$ GeV.}
In $Pb+Pb$ collisions we would expect a similar effect. However
one can see a definite (upto 40\%) overestimate in the calculation. 
The same result was obtained in the LO analysis (see Ref.~\cite{zfpbl02}).
The deviation of the data from the pQCD prediction 
indicates the appearance of a new collective effect in the $Pb+Pb$ collision.
One candidate for this deviation is ``jet quenching'', the induced
energy loss of high energy jets penetrating hot dense matter.
This effect can be seen clearly at RHIC energy (see 
the data in Refs.~\cite{PHENIX,STAR},
and an explanation in Ref.~\cite{QM01}).
Since jet quenching is strongly increasing with energy,
its moderate appearance at CERN SPS in central $Pb+Pb$ collisions
is acceptable.

\section{Summary and conclusion}
\label{sec_sum}

We presented a pQCD based parton model calculation 
in next-to-leading order powered by a jet-level $K$ factor. 
The intrinsic transverse momentum 
distribution of the partons inside nucleons was 
included into the model.
The Gaussian width of this distribution in $pA$ and $AA$ collisions
is controlled 
by two terms: the free $pp$ value $\k2av_{pp}$, 
fixed by several different experiments, 
and a nuclear part, which gives extra enhancement due to semihard 
collisions. (Note that at NLO level 
the factorization scale was chosen to be $Q=p_{T,jet}$.)  
While the nuclear part controls the height of the Cronin peak,
the $\k2av_{pp}$ part determines its position.

We analyzed
CERN SPS data and found that in $Pb+Pb$ collisions the experimental
data are  below the pQCD prediction by $30-40$ \%.
This is similar to what was seen
in LO calculations\cite{zfpbl02}. 
With the present choice of scale and NLO approximation,
the overestimation  
appears to be independent
of the order of the pQCD calculation and indicates some
collective nuclear effect. One candidate is the induced jet energy
loss in hot dense matter ("jet-quenching"), which becomes significant
in central $Pb+Pb$ collisions.
 
\vspace*{4 mm}

\section*{Acknowledgments}
\label{sec_ack}

This work was supported in part by  U.S. DOE grant DE-FG02-86ER40251, 
NSF grant INT-0000211, FKFP220/2000 and Hungarian grants OTKA-T032796 
and OTKA-T034842. Supercomputer time provided by BCPL in Bergen, 
Norway and the European Community $-$ Access to Research Infrastructure 
action of the Improving Human Potential Programme is gratefully 
acknowledged. 


\vfill\eject
\end{document}